\documentclass[aps,prd,onecolumn,groupedaddress,showpacs,nofootinbib,amssymb
]{revtex4}
\usepackage{graphicx,bm}
\usepackage[english]{babel}
\usepackage{amsmath}
\usepackage{amssymb}
\usepackage{amsfonts}

\begin{document}

\def\cL{{\cal L}}
\def\be{\begin{equation}}
\def\ee{\end{equation}}
\def\bea{\begin{eqnarray}}
\def\eea{\end{eqnarray}}
\def\beq{\begin{eqnarray}}
\def\eeq{\end{eqnarray}}
\def\tr{{\rm tr}\, }
\def\nn{\nonumber\\}
\def\e{{\rm e}}


\title{Black hole and de Sitter solutions in a covariant renormalizable field theory of gravity}

\author{G. Cognola$^{1}$, E. Elizalde$^{2}$, L. Sebastiani$^1$ and S. Zerbini$^1$}
\affiliation{ \mbox{} \\
$^1$Department of Physics, Trento University and \\ Gruppo Collegato di Trento  Sezione INFN di Padova, Italy
\\
$^2$Consejo Superior de Investigaciones Cient\'{\i}ficas \\
Institut de Ci\`{e}ncies de l'Espai (IEEC-CSIC) \\
Campus UAB, Facultat de Ci\`{e}ncies \\ Torre C5-Par-2a pl,  E-08193
Bellaterra (Barcelona), Spain\\
}

\begin{abstract}
It is shown that Schwarzschild black hole and de Sitter solutions exist as exact solutions of a recently proposed relativistic covariant formulation of (power-counting) renormalizable gravity with a fluid. The formulation without a fluid is also presented here. The stability of the solutions is studied and their corresponding entropies are computed, by using the covariant Wald method. The area law is shown to hold both for the Schwarzschild and for the de Sitter solutions found, confirming that, for the $\beta=1$ case, one is dealing with a minimal modification of GR.

\end{abstract}


\maketitle

\section{Introduction.}

Attempts to quantize gravity have been mainly carried out by considering the perturbations of a flat, Lorentz invariant background and using the principles of General Relativity. By doing this, unavoidable, non-renormalizable divergences
coming from the ultraviolet region in momentum space show up. To escape this severe
difficulty, higher derivative theories have been invoked but there a new problem, namely the unitarity issue, appears (see e.g. \cite{ilb}). Recently, Horava had the idea to directly modify the ultraviolet behavior of the graviton propagator in a Lorentz non-invariant way \cite{Horava:2009uw}, as $1/\left|\bm{k}\right|^{2z}$, with $\bm{k}$ the spatial momenta and $z=2,3$ or higher. This exponent comes from the different scaling properties of the space-time coordinates $\left(\bm{x},t\right)$, as $\bm{x}\to b\bm{x}$, $t\to b^z t$. When $z=3$, the theory appears to be UV power-counting  renormalizable (which leads to conjecture renormalizability). To exhibit the Lorentz non-invariance, terms explicitly breaking Lorentz invariance (or more precisely, the full diffeomorphism invariance) are written down, by treating the temporal and the spatial coordinates differently. The Horawa model has diffeomorphism invariance with respect to the time coordinate $t$ only, while for the spatial coordinates: $\delta x^i=\zeta^i(t,\bm{x})$, $\delta t=f(t)$, with $\zeta^i(t,\bm{x})$ and $f(t)$ arbitrary functions.

In Ref.~\cite{Nojiri:2009th} a Ho\v{r}ava-like gravity model with full diffeomorphism invariance was proposed. There, when considering perturbations from a flat, Lorentz
invariant background, the Lorentz invariance of the propagator was dynamically broken by a non-standard coupling with a perfect fluid. The propagator behaved as
$1/{\bm{k}}^{2z}$ with $z=2,3,\cdots$ in the ultraviolet region and the model could be
perturbatively power counting (super-)renormalizable, if $z\geq 3$.
The price to pay for such covariant renormalizability was the presence of an
unknown fluid, which might have a stringy origin but cannot correspond to a usual fluid, like radiation, baryons, dust, or the like. The model could be consistently constructed when the equation of state (EoS) parameter $w\neq -1\, ,1/3$. For usual particles in the high energy region, the corresponding fluid is relativistic radiation, for which $w\to 1/3$. Actually, the non-relativistic fluid was needed even in the high energy region.
Later, a dust fluid with $w=0$ was constructed for the scalar theory by
introducing a Lagrange-multiplier field, which gives a constraint on the first scalar field \cite{Lim:2010yk,Capoz1}.

More recently \cite{Nojiri:1004}, a fluid with arbitrary constant $w$ from
a scalar field which satisfies a constraint has been constructed. Owing to the constraint, the scalar field is not dynamical and, even in the high energy
region, a non-relativistic fluid could be obtained. Through coupling with
the fluid, a full diffeomorphism invariant Lagrangian results (in fact a class of
such gravitational Lagrangians), which is given completely in terms of fields
variables. It has been demonstrated in \cite{Nojiri:1004} that such theory has
all the good properties of the Lorentz non-invariant gravities, as in the previously mentioned theories (like its conjectured renormalizability \cite{Carloni:2010nx}), while having the advantage of being at the same time a covariant theory. It was also conjectured there, that the spatially-flat FRW cosmology for such covariant field gravity might exhibit accelerating solutions \cite{review}.

In the present paper we do show this to be the case. In particular, we investigate
in the next section the existence of black hole solutions in the lastly mentioned theory of covariant (power-counting) renormalizable gravity and it is shown that Schwarzschild and de Sitter solutions may exist as exact solutions of the same. Also, a covariant model without a fluid will be here presented. In Sect.~III we study possible cosmological applications, by looking explicitly for cosmological solutions, considering therefore a FRW metric and a scalar field depending on time only. The stability of the solutions is analyzed. In the general case, the possible presence of acceleration is also investigated. In Sect.~IV the entropy corresponding to all these solutions is computed \cite{ceoz_jcap}, by using the covariant Wald method. We prove that the area law is satisfied both for the Schwarzschild and for the de Sitter solutions found, confirming that, for $\beta=1$, one is dealing with a minimal modification of GR. Finally, Sect.~V is devoted to conclusions.

\section{Black hole solutions in covariant (power-counting) renormalizable gravity.}

To start, let us briefly review the covariant (power-counting) renormalizable gravity of Ref.~\cite{Nojiri:2009th}. It is described by the action
\be
\label{Hrv1}
I =\frac{1}{16 \pi G}\,\int d^4x\,\sqrt{-g}\, \left\{R-2\Lambda
-\alpha\left[\left(R^{ij}-\frac\beta2\,Rg^{ij}\right)\nabla_i\phi\nabla_j\phi\right]^n
    -\lambda\,\left(\frac12\,g^{ij}\nabla_i\phi\nabla_j\phi+U_0\right)\right\}\ ,
\ee
where $G$ is Newton's constant, $R$ and $R_{ij}$ are the scalar curvature and Ricci tensor, respectively, $\phi$ is a cosmological scalar field,
$\lambda$ a Lagrangian multiplier, $\alpha,\beta,\Lambda,U_0$ are arbitrary constants
and, finally, $n\geq1$ is an arbitrary number.

Variation of the action with respect to $\lambda$ gives the constraint
\be\label{const}
g^{ij}\nabla_i\phi\nabla_j\phi=-2U_0\,,
\ee
while the field equations for the scalar field read
\bea\label{PHI}
0&=&\nabla_i\,\left\{\left[2n\alpha\,F^{n-1}\left(R^{ij}
       -\frac\beta2\,Rg^{ij}\right)+\lambda\,g^{ij}\right]\,\nabla_j\phi\right\}
\nonumber\\ &=&\frac1{\sqrt{-g}}\,\,
\partial_i\,\left\{\left[2n\alpha\,F^{n-1}\left(R^{ij}
       -\frac\beta2\,Rg^{ij}\right)+\lambda\,g^{ij}\right]\,\sqrt{-g}\,
                 \partial_j\phi\right\}\,,
\eea
where, for convenience, we have set
\be
F=T_{ij}R^{ij}-\frac\beta2\,RT\,,\qquad T_{ij}=\nabla_i\phi\nabla_j\phi\,,
                            \qquad T=g^{ij}T_{ij}=-2U_0\,.
\ee

The field equations related to the gravitational field have the form
\bea\label{Gij}
G_{ij}+\Lambda g_{ij}+\frac{\alpha}{2}\,F^n\,g_{ij}&=&
      n\alpha F^{n-1}\left[R^k_iT_{kj}+R^k_jT_{ki}-\frac\beta2\left(TR_{ij}+RT_{ij}\right)\right]
+\frac\lambda2\,T_{ij}
\nn &&\qquad
         +n\alpha\,\left[D_{rsij}(T^{rs}F^{n-1})-\frac\beta2\,D_{ij}(TF^{n-1})\right]
+\Omega^{rs}\,\frac{\delta T_{rs}}{\delta g^{ij}}\,,
\eea
where $\Omega_{rs}$ is a tensor which will play no role in the following, and we have introduced the differential operators
\be
D_{ij}=g_{ij}\Box-\frac12\,(\nabla_i\nabla_j+\nabla_j\nabla_i)\,,
\ee
\be
D_{rsij}=\frac{1}{4}\,\left[(g_{ir}g_{js}+g_{jr}g_{is})\Box
                           +g_{ij}(\nabla_r\nabla_s+\nabla_s\nabla_r)
                         -(g_{ir}\nabla_s\nabla_j+g_{jr}\nabla_s\nabla_i
                      +g_{is}\nabla_r\nabla_j+g_{js}\nabla_r\nabla_i)\right]\,.
\ee
Note that the field equations in (\ref{Gij}) are valid for an arbitrary, symmetric
``energy-momentum'' tensor $T_{ij}$, but in our particular case such a tensor does not
depend on the metric and so the last term in (\ref{Gij}), depending on $\Omega_{rs}$, drops out.
Now, we look for interesting physical solutions of the field equations above.
\begin{description}
\item{\bf Schwarzschild solution:}
this is the simplest one and can be easily obtained for $\Lambda=0$ and $n>1$.
In fact, in all such cases $R_{ij}=0, \lambda=0$ satisfy all field equations.
The scalar field $\phi$ has to fulfill the constraint (\ref{const}) only.

\item{\bf Einstein-space solutions:} these are generalisations of the previous solutions.
They have the form
\be
R_{ij}=\frac14\,R_0\,g_{ij}\,.
\ee
In such a case
\be
F=\left(\beta-\frac12\right)\,R_0U_0\equiv F_0
\ee
is a constant and, from (\ref{PHI}) and (\ref{Gij}), we get
\be\label{lamb}
g^{ij}\nabla_i\,\left[n\alpha\left(\frac12-\beta\right)\,R_0F_0^{n-1}
                    +\lambda\right]\nabla_j\phi=0\,,
\ee
\be\label{EINS}
\left[\Lambda-\frac{R_0}{4}+\frac{\alpha}{2}\,
               \left(1+\frac{n\beta}{1-2\beta}\right)\,F_0^n\right]\,g_{ij}=
                \frac\lambda2\,T_{ij}
             +n\alpha\,F^{n-1}\left(D_{rsij}\,T^{rs}+\frac{1-\beta}2\,R_0T_{ij}\right)\,.
\ee
We see that non-trivial solutions effectively exist. For example,
if $\lambda$ and $\phi$ satisfy the equations
\be
\lambda=n\alpha\left(\beta-\frac12\right)\,R_0F_0^{n-1}\,,
\ee
\be\label{ExCond}
D_{rsij}\,T^{rs}+\frac14\,R_0T_{ij}=\Sigma g_{ij}\,,
\ee
$\Sigma$ being a constant.
In such case, the curvature can be derived from the algebraic equation
\be
\frac{R_0}4-\Lambda+\alpha\,\left\{n\Sigma+\frac{R_0U_0}4\,
\left[1-(n+2)\beta\right]\right\}\,
\left[\left(\beta-\frac12\right)R_0U_0\right]^{n-1}=0\,.
\ee
Of course this is a solution if the equations (\ref{ExCond}) are compatible
with the constraint (\ref{const}).
In principle, more general solutions with non-constant $\Lambda$ may exist too.
\end{description}

\section{Cosmological applications}
We shall now look for cosmological solutions and thus we start with a FRW metric and
a scalar field which depend on time only. Then, $\phi=\phi(t)$ is
completely determined by the constraint (\ref{const}) and, as a consequence,
the tensor $T_{ij}$ has only one non-vanishing component, namely $T_{00}=\dot\phi^2=2U_0^2$.

Since all quantities depend on time only, the equation in (\ref{PHI}) gives
\be\label{lat}
\lambda
  -n\alpha\left[6\left((\beta-1)\dot H+(2\beta-1)H^2\right)\right]^nU_0^{n-1}
    =\frac{C}{a^3}\,,
\ee
$H(t)=\dot a/a$ being the Hubble parameter and $C$ an arbitrary integration constant.
Moreover, due to the symmetry of the metric in (\ref{Gij}), only two equations are independent.
It is clear that, by choosing $\beta=1$, one has a simplification, namely
\bea
0&=& \Lambda-3H^2+\frac12\,\alpha\,(1-4n)(6U_0H^2)^n+U_0\lambda\,,
\\
0&=& \Lambda-3H^2-2\dot H+\frac12\,\alpha(1-2n)(6U_0H^2)^n
                          +\frac13\,\alpha n(1-2n)\dot H(6U_0)^nH^{2n-1}\,.
\eea
Now, in the latter equations, $\lambda$ can be eliminated by means of (\ref{lat}),
getting in this way the generalized Friedmann equations for the pure gravitational field. We have
\bea
0&=& \Lambda-3H^2+\frac12\,\alpha\,(1-2n)(6U_0H^2)^n-\frac{C}{a^3}\,,
\\
0&=& \Lambda-3H^2-2\dot H+\frac12\,\alpha(1-2n)(6U_0H^2)^n
                          +\frac13\,\alpha n(1-2n)\dot H(6U_0)^nH^{2n-1}\,.
\eea
One easily sees that, in order to get de Sitter solutions, one has to choose
a vanishing integration constant, that is $C=0$. In this way the previous
equations become equivalent and one obtains the Hubble constant $H(t)=H_0$ by solving
\be
\frac12\,\alpha\,(2n-1)(6U_0H_0^2)^n+3H_0^2-\Lambda=0\,.
\ee
On the contrary, choosing $C\neq0$ one gets a second-order
differential equation in the variable $a(t)$. A simple way to get such equation is to make use of the well known minisuperspace approach, which we will briefly describe in the following.

Recall we  are dealing with a FRW space-time, namely
\be
ds^2=-e^{2b(t)}dt^2+a(t)^2d(\vec x)^2\,,
\ee
here $b(t)$ describes the reparametrization invariance of the model, and we assume $\phi$ is a function of time, $t$, only. As a result,  one has
\be
F=K^{ij}\partial_i \phi \partial_j \phi =(R^{ij}-\frac{\beta}{2}Rg^{ij})\partial_i \phi \partial_j \phi=
3 (\dot \phi)^{2}e^{-4 b}\left[(\dot a)^{2}a^{-2}+(\beta-1)(\frac{ \ddot a}{a}-\frac{\dot a \dot b}{a} )\right]\,.
\ee
One can see the particular role played by the dimensionless parameter $\beta$. If one makes the choice $\beta=1$,
namely $K_{ij}=G_{ij}$, the Einstein tensor, the dependence on the acceleration and $\dot b$ drops out. In fact, due precisely to the diffeomorphism invariance of the model, $ G_{00}$ is the Hamiltoniam constraint of GR and the modified
gravitational fluid model becomes very  simple, so that one has  the following simplified minisuperspace action
\be
I=\frac{1}{16\pi G}V \int dt \left[ -6 a (\dot a)^2 e^{-b}-2\Lambda a^3e^b -\alpha 3^ne^{(1-4n)b}(\dot a)^{2n}a^{-2n+3}
(\dot \phi)^{2n}-\lambda a^3 e^b(U_0-e^{-2b}\frac{(\dot \phi)^2}{2})\right]\,,
\ee
In this case,  one has two Lagrange multipliers $\lambda$ and $b$,
the first one implements the constraint
\be
U_0=e^{-2b}\frac{(\dot \phi)^2}{2} \,,
\label{c1}
\ee
while the second gives the Hamilonian constraint of our covariant model. After the variation,  one has to take $b=0$.
The other two Lagrangian coordinates are $\phi$ and $a$, and one has the corresponding equations of motion.
Let us continue with the equation of motion associated with $b$.
On shell, one has
\be
6H^2-\alpha(1-4n)(6U_0)^nH^{2n}-2\Lambda=2\lambda U_0\,.
\label{b}
\ee
On the other hand, since the Lagrangian does not depend on $\phi$, the associated equation of motion reads
\be
C=\frac{\partial L}{\partial \dot \phi}
\ee
where $C$ is a constant of integration. On shell,
\be
-2n\alpha(6U_0)^nH^{2n}+2\lambda U_0=\frac{C \sqrt{2 U_0}}{a^3}\,.
\label{p}
\ee
Making use of the two last equations, we arrive at
\be
6H^2-\alpha(1-2n)(6U_0)^nH^{2n}-2\Lambda=\frac{C \sqrt{2 U_0}}{a^3}\,.
\label{nol}
\ee
Finally, the last equation of motion is the one associated with $a$. It reads
\be
(6H^2-\alpha(1-2n)(6U_0)^nH^{2n}-2\Lambda)=-(4+\alpha\frac{2n}{3}(2n-1)(6U_0)^nH^{2n-2} )\dot H\,.
\label{a}
\ee
And making use of above equations, we also have
\be
\frac{C \sqrt{2 U_0}}{a^3}=-(\alpha\frac{2n}{3}(2n-1)(6U_0)^nH^{2n-2}+4 )\dot H\,.
\label{a1}
\ee

Some remarks are here in order. The equations we have obtained are identical to the ones coming directly from the equations of motion. In particular, as in General Relativity, the equation of motion associated with $a$ is not an independent one, since it can be obtained by taking the derivative with respect to $t$ of the other equations, and de Sitter solution corresponds to the choice $C=0$. In this case, Eq.~(\ref{a1})
is satisfied, and we have
\be
6H^2+\alpha(1-2n)(6U_0)^nH^{2n}-2\Lambda=0\,.
\label{no2}
\ee
One needs to look for positive $x=H^2$ solutions with $\alpha >0$, a necessary condition in order to have a correct non linear
graviton dispersion relation \cite{Nojiri:2009th}. With regard to this issue, let us consider the simplest non trivial case,
namely $n=2$. One has as a solution
\be
H^2=\frac{-1+ \sqrt{1+24\alpha U_0^2\Lambda}}{36\alpha U_0^2}\,.
\ee
Note that, for $\Lambda=0$, the de Sitter solution  exists only for $\alpha<0$, which would correspond to an unusual dispersion relation for the graviton.

The stability of all de Sitter solutions is not difficult to study. In fact taking the first variation of Eq.~(\ref{a}) around $H=H_0$, one obtains
\be
\frac{d \,\delta H}{dt}=-3 H_0 \delta H\,.
\ee
As a consequence, all the de Sitter solutions are stable.

Let us investigate the case when $C$ is non-vanishing. In this case a de Sitter solution does not exist. Then, we may take $\Lambda=0$. First, let us study the
model with  $n=2$. In this case, with $\alpha >0$, one has the differential equation
\be
\frac{d H}{d t}=-\frac{3}{2}\frac{ H^2+18\alpha U_0^2 H^4}{1+36\alpha U_0^2 H^2}\,.
\ee
Separating variables, one gets
\be
\frac{1}{H(t)}-6U_0\sqrt{\frac{\alpha}{2}}\mbox{arctan}(6U_0\sqrt{\frac{\alpha}{2}}H(t))
=\frac{3}{2}t+B\,,
\label{generalsolution}
\ee
where $B$ is an integration constant. The solution is given in an implicit way only.  However, even then it is easy to show that the model is protected against future-time singularities. In fact, let us look for solutions in which the Hubble parameter is
expressed as
\begin{equation}
H=\frac{h}{(t_{b}-t)^{a}}\label{Hsingular}\,,
\end{equation}
where $h$ and $t_{b}$ are positive constants and $t<t_{b}$, since we are investigating the future of an  expanding universe.
The exponent  $a$ is a positive constant or either a negative non-integer number, so that, when $t$ is close to $t_{b}$, $H$ or some derivative of $H$, and therefore the curvature, become singular. When $ a>0$, the arctangent in Eq.~(\ref{generalsolution}) tends to a constant and the sign of the first leading term on the left hand side is inconsistent with the sign of the right hand side. As a consequence, no singular future solution can exist.  Moreover, when $ a<0$, the lhs of Eq.~(\ref{generalsolution}) diverges.

In the general case, we can investigate the possible presence of acceleration. In fact, with  $\Lambda=0$, one has
\be
\frac{\dot H}{H^2}+1=\frac{1}{(2+\alpha\frac{n}{3}(2n-1)(6U_0)^nH^{2n-2} }\left[ -1+\alpha(2n-1)(2n-3)6^{n-1}U_0^nH^{2n-2} \right]
\ee
As a result, in the case when matter can be neglected, one may have acceleration as long as
\be
H^{2n-2}>\frac{1}{\alpha(2n-1)(2n-3)6^{n-1}U_0^n }\,.
\ee
In particular, for $n=2$ this condition becomes
\be
H^2>\frac{1}{18 \alpha U_0^2}\,.
\ee

Coming back to the general model, it turns out that for $\beta \neq 1$ calculations are much more involved, since $ \ddot a$ is present in the Lagrangian, and the model
becomes a higher-derivative system in the sense of Ostrogradsky. However, we may carry out a direct calculation, which shows that a dS solution is not possible there.

\section{Entropy calculation}

It is of interest to evaluate the black hole entropy associated with the different solutions we have discussed. Since we are dealing with a covariant theory, we can make use of the Noether charge Wald methods. The Wald formula reads \cite{wald,visser}
\be
S=-2\pi\int\frac{\partial L}{\partial R_{ijrs}}\varepsilon_{ij}\varepsilon_{rs}\,dA\,,
\ee
where the integral is over the two-dimensional horizon, a spherical surface, and
$\varepsilon^{ij}$ is the binormal tensor to this surface,
normalized as $\varepsilon_{ij}\varepsilon^{ij}=-2$. A direct evaluation yields
(cf.~with Ref.~\cite{ceoz_jcap})
\be
\frac{\partial L}{\partial R_{ijrs}}\varepsilon_{ij}\varepsilon_{rs}=
    \frac{1}{16\pi G}\,\left[(
\varepsilon_{ij}\varepsilon^{ij}
     -n\alpha F^{n-1} \frac{\partial F}{\partial R_{ijrs}}\varepsilon_{ij}\varepsilon_{rs}\right]\,.
\ee
The first term is the GR contribution, while the other one is due to the modification of GR in the considered model. However, in the case of the Schwarzschild solution one has $F=0$. Thus,
\be
S=\frac{A_H}{4G}\,,
\ee
where $A_H=4\pi r_H^2$. As a consequence, in this modified gravity model, the entropy of  the Schwarzschild black hole satisfies the usual Area Law.

Let us now consider the dS solution we have found for $n=2$. The simplest way to perform the calculation is to make use of the static gauge, namely
\be
ds^2=-V(\rho) dt^2_s+\frac{d \rho^2}{V(\rho)}+\rho^2 d\Omega^2\,,
\ee
being  $V(\rho)=1-H_0^2\rho^2 $. This static form of the dS metric can be obtained from the FRW by the coordinate transformation
\be
\rho=r e^{H_0 t}\,, \quad t=t_s+\frac{1}{2 H_0}\ln V(\rho)\,.
\ee
 The solution corresponding to the scalar fluid reads
\be
\phi(t_s,\rho)=\sqrt{2 U_0}[t_s+\frac{1}{2 H_0}\ln V(\rho)]\,.
\ee
The relevant scalar quantity to be evaluated is
\be
\frac{\partial F}{\partial R_{ijrs}}\varepsilon_{ij}\varepsilon_{rs}=-
2U_0+\varepsilon_{ij}\varepsilon_{rs}\partial^i \phi \partial^r \phi g^{js}\,.
\ee
In general, the binormal tensor is given by  $\varepsilon_{ij}=v_i u_j-v_j u_i$ and, in a static gauge, it is easy to show that one may choose  $v_i=(\sqrt{V},0,0,0)$ and $u_i=(0,\frac{1}{\sqrt{V},0,0})$. A direct calculation yields
\be
\varepsilon_{ij}\varepsilon_{rs}\partial^i \phi \partial^r \phi g^{js}=2 U_0 \,.
\ee
Thus, the Area Law is also satisfied for the de Sitter solution we have found, confirming that, for $\beta=1$, we are dealing with a minimal modification of GR.

\section{Conclusions}

In Ref.~\cite{Nojiri:1004}, a fluid with arbitrary constant $w$ from
a scalar field which satisfies a constraint was constructed in a way that, through the coupling with the fluid, a full diffeomorphism invariant Lagrangian was obtained, given completely in terms of fields variables. This theory was shown to have
all the good properties of previous Lorentz non-invariant gravities, like its conjectured renormalizability (coming from the corresponding property that holds at the power-counting level), with the additional advantage of being at the same time a covariant theory. It was conjectured that the spatially-flat FRW cosmology associated to this covariant field gravity might have accelerating solutions. Here we have shown this to be in fact the case. The formulation without a fluid has also been presented. In particular, we have demonstrated the existence of Schwarzschild black hole and of de Sitter solutions, under some conditions, as exact solutions of the theory. We have considered possible cosmological applications, by looking for cosmological solutions, where the FRW metric and the scalar field depend on time only. The stability of the de Sitter solutions has been studied and, in the general case, the possible presence of acceleration has been investigated too. Finally, the entropies corresponding to the different solutions have been obtained, to prove that the area law is satisfied both for the Schwarzschild and for the de Sitter solutions found, thereby confirming that, for $\beta=1$, one is actually dealing with a minimal modification of GR.
\medskip

\noindent {\bf Acknowledgments.} We thank Sergei Odintsov for useful discussions. This research has been carried out in the framework of the Cooperation Agreement INFN(Italy)-DGCYT(Spain). It has also been supported in part by MEC (Spain) project FIS2006-02842 and by AGAUR (Generalitat de Ca\-ta\-lu\-nya), contract 2009SGR-994.

\end{document}